# Generating and Visualizing Trace Link Explanations

Yalin Liu, Jinfeng Lin, Oghenemaro Anuyah, Ronald Metoyer, Jane Cleland-Huang
University of Notre Dame
Notre Dame, IN
yliu26,jlin6,oanuyah,rmetoyer,JaneHuang@nd.edu

## ABSTRACT

Recent breakthroughs in deep-learning (DL) approaches have resulted in the dynamic generation of trace links that are far more accurate than was previously possible. However, DL-generated links lack clear explanations, and therefore non-experts in the domain can find it difficult to understand the underlying semantics of the link, making it hard for them to evaluate the link's correctness or suitability for a specific software engineering task. In this paper we present a novel NLP pipeline for generating and visualizing trace link explanations. Our approach identifies domain-specific concepts, retrieves a corpus of concept-related sentences, mines concept definitions and usage examples, and identifies relations between cross-artifact concepts in order to explain the links. It applies a post-processing step to prioritize the most likely acronyms and definitions and to eliminate non-relevant ones. We evaluate our approach using project artifacts from three different domains of interstellar telescopes, positive train control, and electronic healthcare systems, and then report coverage, correctness, and potential utility of the generated definitions. We design and utilize an explanation interface which leverages concept definitions and relations to visualize and explain trace link rationales, and we report results from a user study that was conducted to evaluate the effectiveness of the explanation interface. Results show that the explanations presented in the interface helped non-experts to understand the underlying semantics of a trace link and improved their ability to vet the correctness of the link.

## 1 INTRODUCTION

Software traceability establishes connections between related artifacts, and then utilizes those links to support numerous software engineering tasks such as safety assurance, impact analysis, and compliance verification [12]. However, given the non-trivial cost and effort of manually creating trace links, researchers have vested significant effort into automating the process using information retrieval (IR) [7, 29, 39], machine learning (ML) [11], and more recently, deep-learning techniques (DL) [22, 34]. In general, a project stakeholder will issue a trace query, generate a set of links, and inspect the resulting links to accept or reject individual results either as a standalone vetting activity [28, 43] or at point-of-use. Trace links generated using IR and ML approaches are often easy to analyze, but tend to deliver relatively low accuracy on large industrial datasets [37, 39]. However, recent advances in DL tracing techniques have returned far higher-degrees of accuracy. For example, in tracing from requirements to code, Lin et al., showed that their TraceBERT approach, which leveraged pretrained BERT models and applied multi-staged fine-tuning, delivered highly accurate trace results for three large, open-source systems achieving Mean Average Precision (MAP) scores greater than 0.86 across several large datasets [34]. Unfortunately, DL-generated trace links can be difficult to interpret without supporting explanations of their underlying semantics.

For example consider a requirement stating that 'The robot shall move to the next position in the order specified by the task plan', and a corresponding design definition that 'The RCU shall publish an AckermannDriveStamped message to the robot's control topic'. Despite having no meaningful common terms the artifacts are linked because the design solution contributes towards satisfying the requirement. An analysis of the concepts shows that *AckermannDriveStamped messages* are closely associated with movements (i.e., carry velocity, angles, and timestamps) and that task plans often involve movement. A domain expert, in this case, someone familiar with the Robotic Operating System (ROS), could likely inspect the two artifacts, apply their innate knowledge of the domain, and intuitively understand the connection between the AckermannDriveStamped messages and the robot's movement. However, someone lacking domain expertise or knowledge of the specific project may have difficulty understanding the underlying concepts and connecting the two artifacts. We therefore believe that explanations are useful across a range of expertise levels including non-experts (e.g., students) and those with partial domain knowledge (e.g., onboarding team members).

This paper therefore addresses the challenge of *trace link explainability*, defined as the ability to explain why two artifacts are related to each other. *Trace explanations* can include both textual information as well as visualizations, and are designed to facilitate reasoning and understanding of semantic relations. They are particularly important when analysts or trace link users lack domain expertise to independently understand the underlying semantics of a trace link, especially as it has been shown that non-experts often mistakingly discard correct links during the link vetting process [28]. Our work addresses a current gap in the literature, as previous studies have focused on trace link accuracy [39, 54], maintainability [41] and efficiency[33], while overlooking the importance of explainability. In prior work, Dick [30], proposed the use of trace rationales in which link creators or vetters would document the rationales behind a link. However, manually annotating links with rationales requires nontrivial effort, thereby increasing the overall cost and effort of creating links. Second, as previously stated, DL approaches often produce links with underlying rationales that are obscure to non-experts. We therefore propose an approach for explaining trace links that aim to automatically generate explanations through mining, extracting, and learning rationales from external knowledge sources.

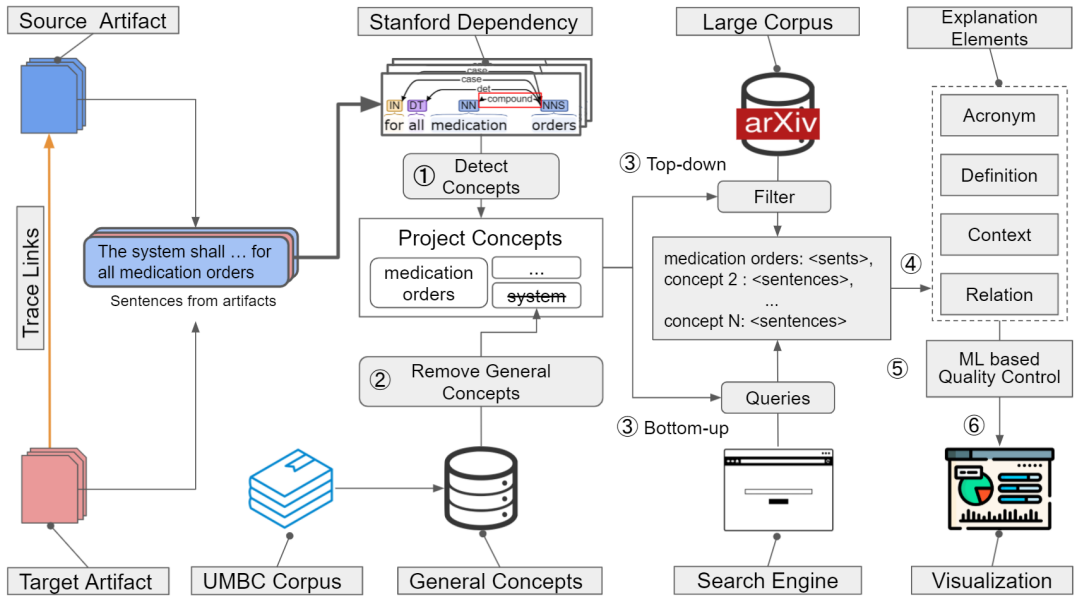

Figure 1: Workflow of our method for extracting the explanation elements from scratch.

Our NLP pipeline executes the following automated steps to generate a link explanation. First, it extracts domain-specific concepts from the artifacts, and then uses these concepts within search queries to retrieve the broader context of each concept from diverse knowledge sources. The resulting dataset constitutes a *context corpus* for the target project. Next, it applies Natural Language Processing (NLP) techniques to extract various forms of structured knowledge from the context corpus, and finally, incorporates this structured knowledge into a trace link explanation that includes both textual descriptions and visualization techniques. We evaluate our approach against three industrial datasets, reporting accuracy and coverage metrics. Further, we conduct a controlled user study and report results showing that utilizing the generated descriptions as rationales within a trace link explanation interface helps non-experts to understand artifact and link semantics and to perform the trace link vetting task more effectively.

Our work makes three primary contributions. First, we propose, implement, and evaluate an NLP pipeline for automatically identifying domain specific concepts and mining acronym expansions, definitions, and contextualized usage examples. Second, we address the challenging problem of data sparsity for specific project domains by integrating both top-down and bottom-up data mining techniques so that we can adapt our approach to different domains with different data sources. Finally, we evaluate the effectiveness of our approach through designing and evaluating an explanation interface with non-expert users. The remainder of this paper is organized as follows. Section 2 provides additional background information. Sections 3 and 4 describe the three datasets used in our study and present the NLP pipeline used to generate each part of our trace link explanation. Section 5 takes a quantitative look at each technique and its utility across three datasets, while Section 6 describes our design of the explanation interface and describes the controlled user-study that was conducted to evaluate its effectiveness. Finally Sections 7 to 9 discuss threats to validity, related work and present conclusions.

## 2 TRACE LINK EXPLANATIONS

As previously explained, we seek to generate explanations that explain the rationale for trace links in a way that is useful for non-experts, as they are the users who experience the greatest difficulty in evaluating the correctness of a link.

### 2.1 The Semantic Gap

Various types of artifacts, such as requirements and design, are often written using different and potentially mismatched terminology. This mismatch can make it challenging for non-experts in the domain to understand why two artifacts are connected by a trace link. A few researchers have explored ways to bridge this gap. Guo et al. proposed a technique for generating trace link rationales [25]; however, their explanations were deeply coupled with heuristics embedded in their underlying trace-link generation algorithms [26]. Liu et al. [36] used the generalized vector space model to improve the quality of generated trace links, and used the HiGrowth algorithm [59] to construct a hierarchical model in which concepts were linked through synonyms, acronyms, ancestors, and siblings; however, they did not consider generating trace link explanations. In this paper, we explore these, and additional techniques, with a focus on explaining the semantics of each individual artifact and the conceptual relationships across pairs of linked artifacts.

### 2.2 Artifact Semantics

Trace link explanations should first provide descriptions of concepts within individual artifacts. We identified three important aspects of *artifact explanations* for enhancing the understandability of technical artifacts. These included (1) acronym expansions, (2) concept definitions, and (3) contextualized usage examples. Building upon our previous example we could expand the internal project acronym RCU to "Robotic Control Unit", provide a definition for an *AckermannDriveStamped message* as 'Time stamped drive command for robots with Ackermann steering', and provide an example context for the use of the term *AckermannDriveStamped message*.

Table 1: Software project datasets used for explanatory and evaluation purposes are drawn from three distinct domains

| Name | Domain | Source | | Target | | Trace Links | Available |
|---|---|---|---|---|---|---|---|
| | | Count | Artifact Description | Count | Artifact Description | | |
| CCHIT | Electronic Health Records | 117 | World Vista Requirements | 588 | CCHIT Regulations | 1065 | http://coest.org |
| CM1 | NASA - Telescope | 54 | Low-level requirements | 23 | High-level requirements | 46 | http://coest.org |
| PTC | Positive Train Control | 263 | Subsystem requirements | 964 | System requirements | 583 | Proprietary |

### 2.3 Link Semantics

The explanation also needs to describe the semantic relationship between two linked artifacts by identifying related concepts across source and target artifacts. We explore two primary ways in which concept-to-concept associations could be explained. First, as a triplet written as $< c_i, v, c_j >$, where the concepts $c_i$ and $c_j$ are connected with a descriptive phrase $v$ to indicate their correlation [38]. As Liu et al. [36] demonstrated, there are multiple ways in which two semantically related concepts can be connected over a relation path. Examples might include (*'AckermannDriveStamped msg', 'published to', 'teleop topic'*) or (*'teleop', 'controls','robot movement'*). While we would ideally use natural language [56] to generate explanations, dynamically constructing clear and concise sentences is a difficult challenge, and therefore we focus this paper on the initial challenge of discovering meaningful triplets that explain trace link relations.

### 2.4 Proposed Solution

Fig. 1 provides a high-level overview of our approach. In Step 1, we analyze project-level artifacts to extract domain-specific concepts – focusing on noun phrases. This step produces thousands of candidate phrases, including both domain-specific and commonly used phrases. Step 2 then filters the list of concepts identified in the project artifacts to remove general concepts (e.g., data structure, user interface). The remaining concepts become the targets of our explanations – first for individual artifacts, and second as part of the trace link explanations. In step 3, we retrieve a domain corpus of sentences containing these concepts, exploring two techniques based on top-down filtering of a large domain corpus, and bottom-up search, driven by the project-specific concepts. This produces a large corpus of sentences – each of which includes at least one targeted domain-specific concept. Step 4 applies a variety of NLP techniques to expand acronyms, generate definitions, discover context, and to build relation triplets – all of which are needed in the explanation. In Step 5, we build a machine learning quality control model which automatically filters non-relevant sentences to improve the accuracy of our explanations. Finally, in Step 6, these explanatory elements are visualized and presented to the user within the explanation interface.

## 3 PROJECT DOMAINS: DATASETS

Throughout the remainder of this paper we focus explanations and experimental analysis on three target software engineering domains of electronic health-care, a space telescope, and positive train control (cf. Table 1). The domains were selected for their diversity, availability of project artifacts, and because each one represented a technical domain with specific terminology and jargon. Two datasets are publicly available, whilst one is proprietary and provided by our industrial collaborators.

- **CCHIT** is from the domain of electronic health-care records (EHR) and includes trace links between 117 requirements from the US Veteran's WorldVista healthcare system (e.g., 'The system shall allow event-delay capability for pre-admission, discharge, and transfer orders'), and 588 regulatory requirements specified by the USA Certification Commission for Health Information Technology (CCHIT) (e.g., 'The system shall provide the ability to send a query for medication history to PBM or pharmacy to capture and display medication list from the EHR').

- **CM1** includes 54 low-level requirements for a NASA spacecraft telescope (e.g., 'The TMALI CSC serves as an intermediate manager of EVENT data supplied by the DCI Driver...'), traced to 23 higher level ones (e.g., 'The DPU-TMALI shall be capable of making data available from the DCI to DPU-DPA. DPU-TMALI will populate a ring buffer...'). This dataset is quite small and the project contains obscure technical jargon with limited online documentation.

- **PTC** is from the domain of Positive Train Control and is provided by our industry collaborator. It traces 263 subsystem requirements to 964 system requirements. We cannot provide examples due to the proprietary nature of this dataset.

## 4 MINING EXPLANATION ELEMENTS

The concept detection step is designed to extract high-quality phrases, representing key domain concepts, from a text corpus. Several researchers have shown the benefits of a phrase-based approach based on information retrieval, taxonomy construction and topic modeling [16, 20, 21, 52]. For example, Liu et al. [35] proposed a technique that integrates phrase mining techniques with phrasal segmentation, and argued that their approach outperforms many other approaches. It starts by identifying the most common n-grams and then applies quality criteria to remove low-quality, less common concepts; but in doing so it removes potentially obscure technical phrases that are of particular relevance for explaining hard-to-understand trace links. For example, the previously discussed 'AckermannDriveStamped message' would be unlikely to survive the filtering process. The approach requires users to provide a small set of example phrases for training purposes.

Shang [52] proposed *Autophrase* which used phrases from section titles in repositories such as wikipedia, to train their model to extract concept phrases composed of nouns, adjectives, and adverbs. As a precursor to the work we present in this paper, we conducted an initial series of experiments using Autophrase trained on Wikipedia to produce a huge concept list covering a wide range of domains. However, we found that Autophrase did not perform well in our targeted software engineering domains, primarily because Wikipedia lacked sufficient training data for our domains. Autophrase not only overlooked important concepts but also extracted phrases with incorrect grammar and/or redundant adjectives. We concluded that a key obstacle in using ML based models to identify core concepts for trace link explanations is the lack of sufficient training data for specific software engineering project domains.

## 4.1 Adopting a Dependency Analysis Approach

Previous research has noted the importance of noun-phrases in the creation and comprehension of trace links [42, 60, 61]. Given the issues we encountered in extracting meaningful domain concepts using ML techniques, coupled with the importance of noun phrases, we opted to leverage Stanford Dependencies (SD) analysis [13, 14] and focused upon detecting meaningful noun phrases as domain concepts. Stanford Dependency analysis identifies direct relations between tokens within a sentence by categorizing their relationships into pre-defined types. This technique is a rule-based approach built using a pre-trained phrase structure grammar parser. The "compound" relations in SD refer to a noun compound modifier of an NP POS-tag that is used to annotate the head noun in a noun phrase [15]. As illustrated in Fig. 2, we use this type of dependency to determine the boundary of a noun phrase – in this example, applied to a sentence from the EHR domain.

## 4.2 Filtering out General Concepts

Analyzing project artifacts in this way tends to produce a large number of concept phrases – some of which are domain specific phrases, worthy of explanation, while others are commonly occurring phrases with well-understood meaning. As we do not wish to clutter our explanations with superfluous information, we reduce the concept list to include only domain-specific ones. We achieve this through generating a *black list* of general concepts. This is achieved by applying the Stanford Dependency analysis to a massive domain-independent corpus, identifying the most commonly occurring concepts, and storing them as the black list. For purposes of this study, we used the UMBC webBase corpus [27], which was built using web-scraping in 2007 by the Stanford WebBase project. The dataset contains English paragraphs with over three billion words. It is 13GB in compressed tar file format and is 48G when uncompressed. After applying the concept detection on this corpus we obtained 2,614,601 noun concepts. We ranked concepts by their frequency, included concepts that appeared more than 1,000 times in our black list. The black list represented the top 39,504 ( 1.5%) of the detected concepts. For each of our three datasets, we removed any concept found in the black list. Examples of project-specific concepts as well as blacklisted ones are listed in Table 2.

## 4.3 Constructing a Concept Domain Corpus

The next step in our pipeline focuses on building a domain corpus in which each sentence includes at least one project-relevant domain concept. Sentences are mined from either an existing general corpus or through searching the internet for relevant text. However, we observed a huge discrepancy in the amount of text available for different domains, and this likely accounts for the different outcomes we report later in this paper for each of our three projects. Popular domains, such as biomedical and finance, have numerous related white papers and websites describing the domain, and sometimes even a textually rich, well-organized corpus (e.g. NCBI disease corpus [19], Reuters Corpora[31]), collated by domain experts. However, many software projects have no previously collated text corpus, and furthermore, some relatively obscure domains have very limited web presence. To address these dual problems, we apply automated corpus collection techniques to build a domain corpus for each targeted project whilst ensuring that retrieved documents have sufficient affinity to the targeted project. Our objective is to mine a focused corpus covering all concepts in the target project, and we explored both "top-down" and "bottom-up" corpus collection strategies.

*4.3.1 Top-down approach:* The top-down approach starts from a relatively large corpus and operates downwards to identify and retain relevant text, whilst eliminated all other parts. The corpus documents are first tokenized into sentences, and then the Knuth-Morris-Pratt (KMP) [48] string matching algorithm is used to efficiently examine whether the sentence contains at least one of the identified project concepts. Matching sentences are then mapped to their associated concepts. We utilized the ArXiv repository [2] which includes abstracts and the full-text of academic papers across physics, computer science, biology and four other large domains [1], and then downloaded 248GB of plain text data through an API provided by ArXiv for this purpose.

*4.3.2 Bottom-up approach:* The bottom-up approach starts with an empty domain corpus and then gradually adds data by searching external resources using a public search engine (i.e., Bing for this study). Search queries are formulated from the previously extracted project concepts, and retrieved websites are scraped to find

Table 2: Examples of Project and Blacklisted Concepts

| CCHIT | CM1 |
| --- | --- |
| Patient Health Information | EEPROM filesystem |
| HL7 / ASTM Continuity | X-ray sensitive CCD imager |
| HIPAA Risk Assessment | DPU Task Monitoring |
| Cardiovascular Tests | FSW Tasks |
| NCPDP Script | DCI Error Interrupt |

| PTC | BlackList |
| --- | --- |
| Wayside Data | User Interface |
| EMP Protocol | Team Manager |
| OBU Transitions | Family Health |
| Class C Protocol | Network Operator |
| Train Control Functions | Useful Results |

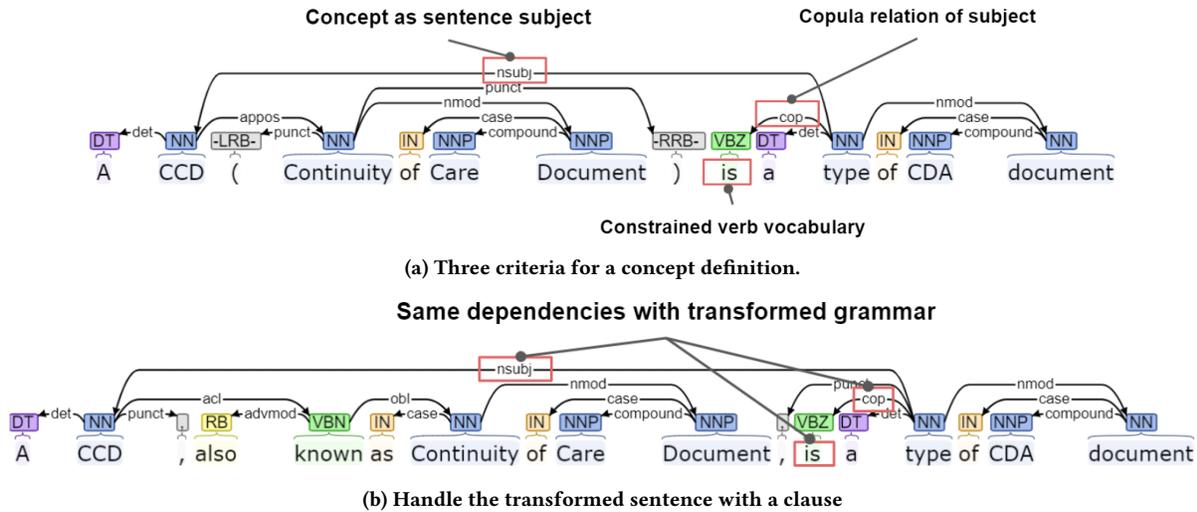

Figure 2: An example of using using Stanford Dependency based rules for concept definition and context extraction. This approach is more robust than pattern matching on complex sentences

sentences containing the targeted concept. We informally experimented with several query templates and found that the template "what is inbody:< *concept* > in < *domain name* >" returned the most consistently relevant results using the Bing search engine, where < *concept* > refers to a project concept that has been automatically detected and < *domain name* > (i.e., Electronic Health Record, Positive Train Control, and NASA) helps the search engine to narrow the scope of the search and to disambiguate similar concepts across different domains. Once results were returned, we apply the same technique as the top-down approach to remove sentences that do not contain at least one project concept.

### 4.4 Extracting Explanation Elements

We then apply a series of NLP techniques to extract the following descriptions and definitions from the constructed corpus.

*4.4.1 Acronym descriptions:* We define a concept as an acronym if all the alphabetic characters in the concept are upper-case. We then utilize the Schwartz-Hearst algorithm [50] to mine the acronyms from the collected corpus. This algorithm leverages pattern matching and heuristic rules (e.g. phrase length) to detect the long and short forms of the acronyms and returns them as mapped pairs. As reported by the authors, the Schwartz-Hearst algorithm achieved 82% recall at precision of 96% when applied to the biomedical domain, outperforming methods previously proposed by Chang et al. [10] and Pustejovsky et al. [47].

*4.4.2 Definitions and Context.* Given the corpus of concept-specific sentences, we utilize Stanford Dependencies heuristic rules to extract definitions and context descriptions. For context extraction, we first check whether a concept is the nominal subject of a sentence and is connected to another word via a "nsubj" dependency. A nominal subject is a noun phrase which is the syntactic subject of a clause [14], and therefore applying this rule ensures that the sentence focuses on the target concept. To identify definitions, we also check the verb connected to the target concept. As the dependencies form a directed relation graph we locate the associated verb by simply searching for verbs that are at most two-hops away from the the target concepts, while simultaneously constraining paths to contain only "nsubj'' and "cop" (copula) dependencies, where the "cop" dependency refers to a relation between a copular verb and its complements. As definitions usually follow a set of known patterns, such as "<concept> is/are/do something", we only select the sentences whose verbs are either "is", "are", or with pos tags of "VBZ" (referring to present tense verbs). We observed the dependencies-based approach to be more robust for handling complex sentences than a simple pattern matching approach, as illustrated in Fig. 2 which shows how rules can be used to extract a definition for "CCD" from sentences with different levels of complexity.

*4.4.3 Filtering non-domain acronyms, definitions, and descriptions:* The automated extraction methods inevitably introduce imprecise results by including sentences and acronyms outside the scope of the target project domains. To address this issue, we developed a deep learning topic model and leveraged it as a binary classifier to identify whether an acronym's long name, a definition, or a context sentence belongs to the target project or not. We trained the model in a weakly supervised manner by utilizing actual artifacts (i.e., requirements, design definitions) from the target project as positive examples and sentences from other sources ( e.g. artifacts from other projects),as negative examples. We used these example to train the model to identify sentences belonging to each target project.

The trained model accepts a tokenized candidate sentence as input, and leverages a BERT-based Language Model (LM) to analyze the semantic meaning of the tokenized sentence. It uses a small Multi-layer Perceptron (MLP) to predict a likelihood score between 0 and 1, representing the affinity between each sentence and the targeted project domain. In this study, we used SciBERT[9] as our language model, which has been pre-trained with massive papers in difference science domains. Based on initial observations of the results, we filtered out all sentences scoring less than 0.5. Further, we ranked all remaining definitions, contextual information, and acronym explanations according to their scores, and in each case, selected the one with the highest score. We evaluated the accuracy achieved with and without this filtering step to test its utility on the entire corpus of retrieved sentences for all three projects and found that accuracy improved from 63.33% to 86.7% for the bottom-up approach and from 50.94% to 66.04% for the top-down approach thereby demonstrating the importance of this part of the pipeline.

*4.4.4 Relation Discovery:* The previous steps have focused on describing concepts found in each individual artifact. In this step we redirect our attention to describing the link itself by applying three techniques for identifying relations between concepts. First, we extract and formulate relations between concepts by exploring the simple *<Subject, Verb, Object>* grammar in the corpus we have collected, by leveraging Stanford dependencies that incorporate verb tokens. More precisely, we use the "nsubj" and "xsubj" tags to find the subject of the verb and "obj" dependencies to identify the object of the verb, and only accept triplets for which the verb represents a hierarchical or equivalency relationship (e.g., includes), as this implies a parent-child relation which can be used to build meaningful cross-artifact relations. For example, the triplet *<navigational information, includes, operational hazards>* could help us to understand why a requirement stating that "The OBU shall transmit navigational information to the back office" is linked to the derived requirement that "The WIU shall detect operational hazards." We created a set of eight seed verbs representing hierarchical relations, and then expanded this set by retrieving four additional synonyms from WordNet [44]. We then combine all of the retrieved triplets into a knowledge graph, in which the subjects and objects of the triplet relations are used as vertices, and their inter-connecting verbs are used as edges. Given two concepts distributed across a pair of linked source and target artifacts, we use Dijkstra's Algorithm [18] to find the shortest path between each candidate concept pair. The path, including its nodes and edges, constitutes a potential explanation for an underlying trace link.

In addition, we consider two concepts as *equivalent* when their lemmatized forms are identical or when one concept is a subsequence of another (e.g. 'TMALI' versus 'TMALI event queue'), in which case we consider the shorter concept to be a semantic abstraction of the longer one.

## 5 QUANTITATIVE ANALYSIS

In our first set of experiments, we investigate the potential usefulness of artifact and trace link explanations by evaluating the correctness of the generated explanation elements and the percentage of artifacts and trace links for which associated elements were mined. For each explanation element and each of the three projects we addressed the following research questions:

**RQ1:** What percentage of the generated explanation elements are correct with respect to the project domain?

**RQ2:** What percentage of the *identified elements* (i.e, acronyms or domain concepts in artifacts) have corresponding correct explanatory elements for use in trace link explanations?

**Table 3: Number of acronyms exist in projects, and the precision and recall score for explaining these acronyms**

|       |         | Top-down  |        | Bottom-up |        |
|-------|---------|-----------|--------|-----------|--------|
|       | Acronym | Precision | Recall | Precision | Recall |
| CCHIT | 109     | 51.79%    | 26.61% | 100.00%   | 11.01% |
| CM1   | 46      | 0.00%     | 0.00%  | 0.00%     | 0.00%  |
| PTC   | 318     | 27.54%    | 8.72%  | 70.00%    | 2.20%  |

To answer these questions, three researchers evaluated the correctness of the generated acronyms, definitions, and context descriptions. In some cases, our domain knowledge was sufficient to directly determine whether a concept was correct; however, in other cases, we reviewed documentation manuals and white papers to discover or confirm the correct meaning of the concept.

### 5.1 Acronym Evaluation

Table 3 reports results for retrieving acronym descriptions for all three projects. The top-down approach generated descriptions for 57, 1, and 69 acronyms at precisions of ~52%, 0%, and ~28% for for CCHIT, CM1, and PTC respectively. This compares to only 15 and 11 acronym descriptions at 100% and ~64% precision for CCHIT and PTC, and no acronyms found for CM1. There are two particularly notable observations. First, the bottom-up approach returned fewer, but more accurate results. This approach likely performed better because search queries included more domain-specific context than the top-down approach. Second, the pipeline completely failed to retrieve any correct acronym descriptions for the CM1 project. While early phases of the pipeline produced 14 acronyms the quality filter correctly eliminated 13 of these as the definitions came from different domains. There are several primary reasons for this failure. First, the acronyms in CM1's design specification refer to very low-level architectural components; second, the CM1 domain contains more technical jargon than either CCHIT or PTC; and third, the domain of interstellar satellites has far fewer online descriptions within white papers, websites, or other documents. For all of these reasons, the acronym expansion failed in the CM1 project but produced useful results for CCHIT and PTC. These results contrast clearly with prior claims that the Schwartz-Hearst [50] algorithm returned 96% accuracy; however those prior results focused on the mechanisms for extracting acronym descriptions from a document in which correct descriptions were available.

### 5.2 Definition and Context Evaluation

Tables 4a and 4b report results from performing the manual evaluation for top ranked definitions and contexts for each concept. The top-down approach retrieved 15, 1 and 16 definitions with accuracies of 80%, 100%, and ~44% for CCHIT, CM1, and PTC respectively, while the bottom-up approach retrieved 12, 3, and 1 definitions at accuracies of ~92%, 100%, and 100% respectively. In the case of contextual descriptions the top-down approach retrieved 85, 7, and 88 definitions at accuracies of approximately 65%, 29%, and 19%, while the bottom-up approach retrieved 44, 8, and 21 definitions at accuracies of approximately 87%, 88%, and 90%. Overall, the bottom-up approach generally retrieved fewer but more accurate results. Again we observe low retrieval rates for CM1, with only one definition in the top-down approach and only three in bottom-up; however,

**Table 4: Coverage and accuracy of the extracted sentences as concept definition and context**

|  | Top-down | | | Bottom-up | | |
|---|---|---|---|---|---|---|
|  | Extr. | Correct | Acc. | Extr. | Correct | Acc. |
| CCHIT | 15 | 12 | 80.00% | 12 | 11 | 91.67% |
| CM1 | 1 | 1 | 100.00% | 3 | 3 | 100.00% |
| PTC | 16 | 7 | 43.75% | 1 | 1 | 100.00% |
| Total | 32 | 20 |  | 16 | 15 |  |

(a) Number and accuracy of project concepts which have definitions extracted from corpus

|  | Top-down | | | Bottom-up | | |
|---|---|---|---|---|---|---|
|  | Extr. | Correct | Acc. | Extr. | Correct | Acc. |
| CCHIT | 85 | 55 | 64.71% | 44 | 39 | 86.67% |
| CM1 | 7 | 2 | 28.57% | 8 | 7 | 87.50% |
| PTC | 88 | 17 | 19.32% | 21 | 19 | 90.48% |
| Total | 180 | 74 |  | 73 | 65 |  |

(b) Number and accuracy of project concepts which have context extracted from corpus

these were retrieved at 100% accuracy. The bottom-up approach also underperformed for the PTC dataset, only finding one correct acronym definition. Despite sending 1,780 unique concept queries to the search engine, only 4.3% of them returned sentences with direct concept matches. This compared to 16.2% and 36.2% direct matches for CCHIT and CM1. This coverage problem has been observed by previous researchers. For example, Zeng et al., reported that only about half of the 30,000 terms found in the MESH Medical taxonomy, appeared anywhere in the PubMed database of 30 million paper abstracts [58]. False positives could likely be further reduced by providing a more diverse set of training examples.

### 5.3 Concept Relation Evaluation

To evaluate concept relations, we examined the generated paths for coverage and correctness. Unfortunately, the knowledge graph suffered from a path sparsity issue and therefore few meaningful paths were found between concepts in paired artifacts. The sparsity was primarily caused by limiting verbs to those representing hierarchical relationships, thereby significantly reducing the size of the triplet set. While accepting a broader set of verbs, creates a far larger set of triplets and many more cross-artifact relations, the majority of these paths do not produce meaningful explanations. We therefore opted to favor precision over recall and excluded multi-hop paths generated from the knowledge graph in our explanation interface. However, the heuristic rules for equivalencies and sub-sequences, along with the 1-hop paths retrieved several interesting explanations such as $< ICD-9, used\ for, billing >$, resulting in 9, 26 and 96 concept relation explanations for the three projects.

### 5.4 Leveraging available Project Glossaries

While our bottom-up approach returned fairly accurate results, there were a large number of artifacts for which no supporting definitions were retrieved. We therefore decided to explore the combination of both the bottom-up and top-down technique alongside definitions provided by project-specific glossaries, and subsequently identified and retrieved a glossary for each project. The CCHIT project traces the requirements in WordVista EHR system to the CCHIT regulatory requirements, and we used a glossary from the WordVista project [5] containing 352 acronyms with their expanded names, and 451 concepts with associated definitions and contextual examples. The provided CM1 glossary [3], provides long names for 64 acronyms used in the project. Finally, we derived the PTC glossary from the architecture specification document[4] of Interoperable Train Control Network (ITCnet) released by the Meteorcomm company and containing 44 acronyms with long names and 69 concepts with definitions. We checked for overlap of definitions. Results reported in Fig. 3 bottom-up (green), top-down (red) approaches, and glossaries (purple), show that different project concepts were provided by each of the three techniques. Adding the definitions generated by the bottom-up approach to each of the existing glossaries increased explanation elements by ~195% for CCHIT, ~33% for CM1, and 150% for PTC. Of course actual gains are highly dependent upon the completeness of the baseline glossary. Given the benefits of combining glossary and generated data, and the accuracy of the bottom-up approach, we used these two data sources in the user study described in the next section of this paper.

## 6 TRACE LINK EXPLANATION INTERFACE

We designed and developed an explanation interface and used it to address the following two research questions:

**RQ4:** Does the trace link explanation interface help users to evaluate the similarity between two linked artifacts more effectively than without the explanation?

**RQ5:** Which aspects of the explanation interface are most helpful to users?

### 6.1 Designing the Explanation Interface

We adopted the explanation design framework introduced by Anuyah et al. [8] to guide the design and evaluation of our explanation interface, and engaged in a series of three participatory design sessions that included six members of our team with expertise in UX-Design and/or Software Engineering. Through these sessions we identified Software Engineering tasks, such as *impact analysis* and *compliance verification* that would utilize traceability, and identified multiple types of users. Of these, we focused primarily on people without domain expertise (e.g., project newcomers) and people performing tasks across skill boundaries (e.g., a business analyst tracing from requirements to low-level models or code) where they may be exposed to terminology they are not familiar with.

We then identified a set of tasks that a user would perform using the interface. These included identifying artifact types, analyzing artifact content, judging whether two artifacts are related, and providing feedback on the correctness of the link. We then brainstormed different ways of presenting the link explanations in support of these tasks, and created a set of *low-fidelity prototypes*. Each prototype was explored to understand its strengths and weaknesses with respect to supporting the potential user tasks and leveraging human perceptual capabilities [40, 45]. This activity informed several design decisions, each of which is reflected in the UI presented in 5.

First, our relation extraction and acronym expansion process results in semantic relations among linked concepts. Connection marks use a line to show a pairwise relationship between two items and thus are appropriate for showing the relationship between two concepts [45]. Second, concepts themselves are nominal types and color encodings are particularly effective for nominal data. Additionally, using the same colors for the same concepts across the two artifacts creates a perceptual grouping [45]. Finally, every concept is accompanied by a quantitative "importance value". Size encodings, while not ideal for showing specific values, are appropriate for comparison purposes and thus we chose to encode the importance of a concept using font size [45].

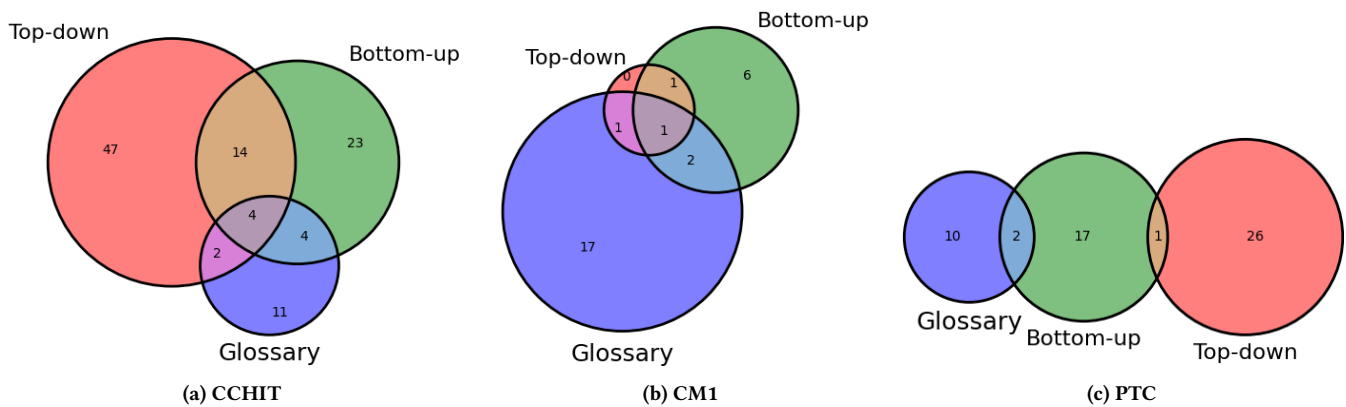

Figure 3: Number of the project concepts explained by project glossaries as well as our top-down and bottom-up approaches

The combined visual elements are intended to provide an overview of the total related elements between the two artifacts, how they are connected, and how strong the connections are. We adopt an overview + detail approach [55] where the user is provided an at-a-glance overview and can obtain details through mouse-over interactions on the concept nodes or concept words themselves to get definitions, or on the connection edges to obtain a semantic relationship description. We iterated through medium-fidelity mockups created in Figma [17] to explore the effectiveness of the visual encodings and interaction strategies. Our final prototype was built as a fully functional web application for use in our evaluation and an interactive demo is available at https://trace-exp-study.github.io/pages.

## 6.2 Study Design

We conducted a controlled user study to evaluate the effectiveness of our trace link explanation interface. We designed our study to have two treatment conditions; showing explanations for the links (TC 1) versus hiding explanations (TC 2). Our goal was to examine the effect that receiving explanations would have for identifying correct and incorrect trace links.

*6.2.1 Participants.* We recruited eight participants with the requirement of having some prior experience working on a software engineering project. We did not exclude participants according to the number of years of experience or their specific role in a project. All of our participants are currently graduate students of a university in the United States and were recruited through email. Participants were tasked to perform trace vetting on 30 links–of which 18 were correct links, while the remaining 12 were incorrect. Each participant evaluated 10 links from each of the three projects. Furthermore, 43% of definitions and acronyms were from the project glossaries, 51% were generated dynamically using the bottom-up approach, and 6% were found in both.

*6.2.2 Procedure.* We began the study by briefing participants about the task that they would perform. Next, we directed participants to our web-based explanation interface using the URL that we provided. Participants were asked to vet links as correct, incorrect, or don't know (if they were unable to confidently make a determination). Participants completed all of the vetting tasks in a random order. In this case, links from the three domains discussed in section 3 were presented randomly. During the session, participants were asked to verbally explain their decision on vetting the links, enabling us to collect both qualitative and quantitative data. The study took about 30-45 minutes to complete.

We exposed each participant to the two treatment conditions, with the aim of understanding the extent to which explanations

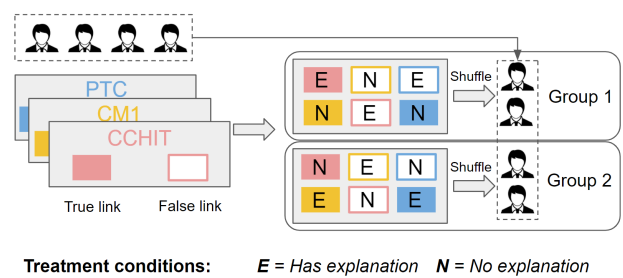

Figure 4: Two experiment groups were formed by sampling an equal number of trace links from each of the three projects. For each link we created a version with an explanation and one without an explanation, and for a given link, Group 1 received the explanation, whilst Group 2 did not. Each group received half the links with explanations and half without, meaning that the two groups received opposite treatments. We then assigned an equal number of study participants to each group. The order of links was randomly shuffled for each participant, who was then asked to evaluate the correctness of each presented link.

Generating and Visualizing Trace Link Explanations

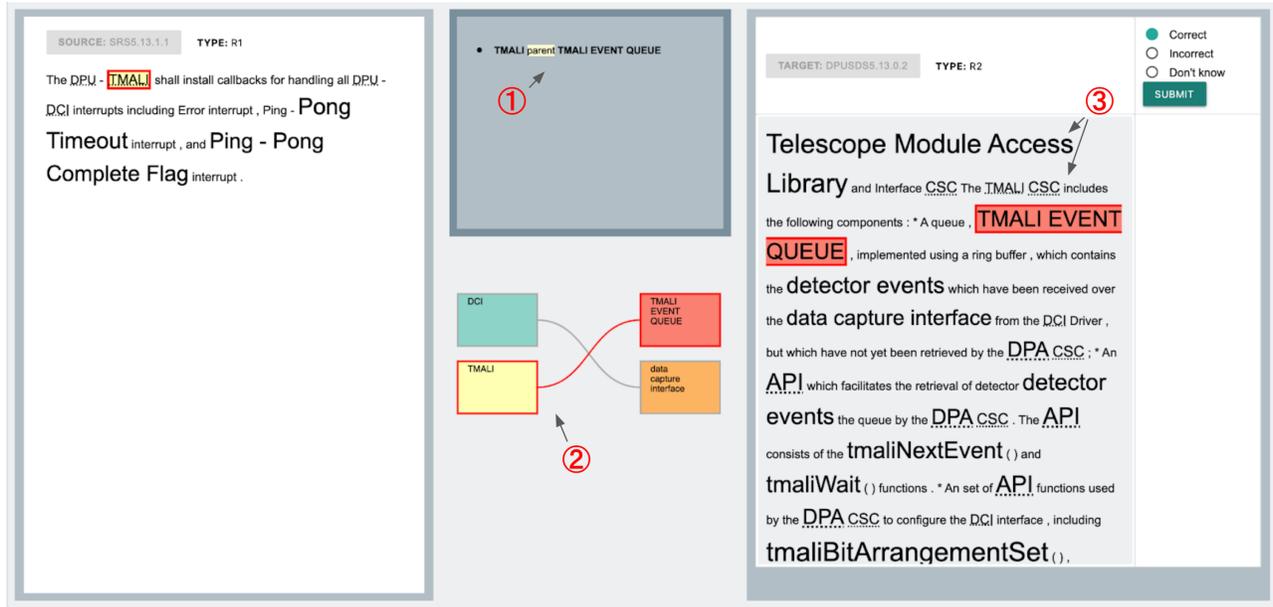

**Figure 5: Snapshot of the trace link explanation user interface. In part 1, we used a text panel to display concept level explanations including long acronym names, definitions, context and inter concept relationships based upon the current cursor (roll-over) position. In part 2, we visualize the links by highlighting the concepts as they appear in the source and target artifacts. In part 3, we enlarge the concepts based on their importance and underscore the ones that have available explanations.**

can guide them in differentiating between correct and incorrect links. To address the potential bias of learning effects, we used a between-subjects experimental design putting each participant in one of two groups, such that participants in one group were exposed to the opposite treatment condition for each of the links, than participants in the second group (see Figure 4). Further, the order in which links were presented to users was randomized for each person.

## 6.3 Results

We examined our data using a combination of quantitative and qualitative measures to understand the impact of our explanation design on trace link vetting.

*6.3.1 Quantitative result outcomes.* Overall, our results show that there was a significant improvement in accurately vetting links when participants were provided with explanations ($\rho$=.01994, $p$ < .05., $d$=0.820852). This finding indicates that the explanations helped to guide our participants for identifying correct and incorrect trace links. However, when we examined the differences for each domain independently, we observed that there were no statistical improvements ($\rho_{PTC}$=.551677, $\rho_{CM1}$=.18026, $\rho_{CCHIT}$=.21684). We attribute this finding to the fact that the data collected in each domain was too small to arrive at a statistical conclusion.

*6.3.2 Qualitative insights.* All participants responded positively to the explanations and expressed how helpful they were for the vetting tasks. None of the participants were experts in the domains under study and were therefore unfamiliar with many of the domain-specific technical terms and acronyms in the artifact content. This resulted in many participants selecting "don't know" when they were not given explanations. P4 stated, *"I don't know what this term means. I want to know what things mean."* P6 noted, *"The explanations helped me to not just easily identify keywords, but better understand what they mean."*

Participants also noted that receiving explanations reduced the mental effort it took for them to comprehend the content of the artifacts, especially when the length was long. Most of the participants struggled to understand the content meaning when presented with long artifacts without explanations, often resulting in the "don't know" response. P8, for example, when presented with a long artifact description, stated *"This artifact requires a lot of thinking. There is too much text and too much function names and abbreviations without explanation."*

Several participants noted that the explanations helped them easily find common terms or keywords in the artifact content. For links without explanations, however, participants often resorted to using the find feature on their browser to identify common terms in the artifact content. Being able to easily identify these common terms in the explanations made it possible for participants to more quickly make a decision. P8 noted that, *"Because drug interactions was highlighted, I identified what was specifically not to be included. Context helped me know what the system is expecting."* P8 also commented about finding keywords in large content artifacts, *"If the text is too huge, highlighting the keywords was helpful. I could easily jump to the important part."* While we did not collect timing information for participants tasks, anecdotally, participants made decisions more quickly when presented with the explanation.

We also observed that the semantic connections between artifact concepts helped our participants in identifying if artifacts were linked. While some participants struggled to determine links when the concept terms did not exactly match, they noted that seeing the relationship path made it possible for them to not only see that artifacts were linked, to examine the relationship more directly to build confidence in their understanding. For example, P3 noted,*"I can easily see that these are not semantically the same."* P7 found it *"...helpful to see overlapping keywords and phrases."*

While the feedback was overwhelmingly positive, participants also noted some areas for improvement. For instance, they noted that some concept terms were too ambiguous to understand, even with definitions. Some of them also mentioned that they needed context for functions and other artifacts that were referenced in the artifact content. P7 stated,*"I don't know what certification is being referred to here. I need some context."*

## 7 THREATS TO VALIDITY

Our study carries a few threats to validity. With respect to *internal validity*, we evaluated our approach on three different systems from diverse domains for which golden answer sets defining correct and incorrect links were already provided. We used domain documentation to evaluate the correctness of the generated definitions and descriptions. With respect to *external validity*, we observed trends across the datasets – such as the observation that the bottom-up approach returned more precise results than the top down one. However, we cannot draw general conclusions based only on three datasets. For example, while we have hypothesized that the reason for CM1's underperformance is that it is a highly scientific system with jargon-filled project artifacts, and therefore general domain documents failed to provide relevant concepts, we cannot categorically support such generalizations at this time.

As with any NLP pipeline, we made numerous design decisions, and whilst we justified our decisions, it is likely that different combinations of techniques would return different results. Finally, with respect to construct validity, we used metrics to show the degree of coverage of the mined acronyms, definitions, and contextual explanations; however, coverage does not measure usefulness. To that end we conducted a small user study, and whilst our participants were graduate students and not currently working in industry, they served as reasonable proxies for project newcomers and other non domain experts working on a project. Nevertheless to more rigorously evaluate whether our approach is useful we need to apply it with real project stakeholders in an actual project context.

## 8 RELATED WORK

In addition to the previously described related work, we briefly summarize other closely related work in concept mining and generating traceability rationales.

*Concept and Relation Mining:* Numerous researchers have focused on techniques for mining concepts and their relations from the web. Angeli et.al[6] proposed dependency analysis based method for triplet relation mining in 2015. Our relation extraction approach is based on the same idea while modify and simplify the heuristic rules to focus on the given noun phrases we detected in project artifacts. Taxonomy expansion algorithms [51, 53] focus on expanding an existing ontology by extracting new concepts from large open corpus and leverage deep learning models to insert the concept into the concept hierarchy. The link explanation task can benefit from these methods when an initial domain concept is available. In the software engineering domain, researchers have proposed or evaluated techniques for ontology building in order to capture key project concepts [23, 32] however, these approaches were used in the trace link creation algorithms and not applied for link explanations.

*Traceability Rationales:* Hull and Dick proposed *Rich Traceability* as a means of explicitly capturing satisfaction arguments between requirements and design, thereby documenting the rationale for a link [30]; however, performing this task manually is very time consuming. Other researchers, such as Balasabrumanian et al. [49] and Zisman et al. [57] proposed specific traceability meta-models describing link semantics; however, while these techniques create semantically typed links, they fail to *explain* the purpose of individual links. Guo et al., proposed a technique that utilized domain-specific knowledge bases to support trace link generation [24, 26], and then augmented the knowledge base with rationale patterns to provide an initial explanation for the link. However, their approach is closely coupled with a heuristic approach to trace link generation, whereas our approach is not dependent upon a specific tracing technique. Finally, a few researchers have explored the use of ontologies to generate semantically meaningful trace links [8, 46].

## 9 CONCLUSION

In this paper we sought to extract acronym descriptions, definitions, contextual examples, and cross-artifact relations by applying an NLP pipeline to project artifacts and large general data corpii. Our goal was to use the generated descriptions within an interface to explain why two artifacts were connected through a trace link. Generating trace link explanations represents a relatively new research challenge designed to support recent advances in DL traceability models which have significantly improved the accuracy of generated trace links, but lack clear explanations.

Our quantitative analysis showed that the generated descriptions and definitions were quite precise and were able to augment manually created project glossaries. While we were not able to generate definitions for all identified project concepts, our user study conducted with non-experts across three different software domains demonstrated that the explanations mined from glossaries and augmented with dynamically retrieved definitions and descriptions aided users in evaluating trace links despite lacking expertise in the domain. In future work we will focus on enhanced techniques for providing more comprehensive coverage of all domain concepts.

We share all of the artifacts for CCHIT and CM1 datasets [1] and the associated code [2] into the public domain to empower other traceability researchers to take up the challenge of generating complete, correct, and meaningful trace link explanations in order to make DL-generated trace links more useful for a broader set of stakeholders.

## 10 ACKNOWLEDGMENT

The work in this paper is primarily funded under the USA National Science Foundation grant CCF-1901059.

---

[1] https://zenodo.org/record/6040328
[2] Github:https://github.com/yliu26/TraceLinkExplanation

Generating and Visualizing Trace Link Explanations